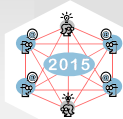

# Regain Control of Growing Dependencies in OMNeT++ Simulations

Raphael Riebl
Technische Hochschule Ingolstadt
Research Centre
Email: raphael.riebl@thi.de

Christian Facchi
Technische Hochschule Ingolstadt
Research Centre
Email: christian.facchi@thi.de

*Abstract*—When designing simulation models, it is favourable to reuse existing models as far as possible to reduce the effort from the first idea to simulation results. Thanks to the OMNeT++ community, there are several toolboxes available covering a wide range of network communication protocols. However, it can be quite a daunting task to handle the build process when multiple existing simulation models need to be combined with custom sources. Project references provided by the OMNeT++ Integrated Development Environment (IDE) are just partly up to the task because it can be barely automated. For this reason, a new approach is presented to build complex simulation models with the help of CMake, which is a wide-spread build tool for C and C++ projects. The resulting toolchain allows to handle dependencies conveniently without need for any changes in upstream projects and also takes special care of OMNeT++ specific aspects.

## I. INTRODUCTION

The desire for using CMake [1] to build OMNeT++ [2] projects has its roots in a true story. When the authors started to use Veins [3] for their simulations, extensions were simply placed within the Veins project tree. Whereas some modifications were considered to be generally useful for Veins and submitted to the upstream project, a growing portion of sources evolved to form a project of its own, which is now called Artery [4]. A number of new modules — first scattered over the Veins source tree — eventually moved to a dedicated source directory, although still within the Veins project. Indeed, Artery does not only depend on Veins, but also on further libraries. Concretely, these are a purely C++ based implementation of Vehicle-to-X (V2X) protocols called Vanetza [5] as well as the quasi-standard Boost [6] libraries. Up to now, Artery managed these additional dependencies by extending Veins' *configure* script. For this purpose it reads a local, user-specific file containing the locations to Boost and Vanetza directories, which can vary from one system to another.

While this approach is working, further extensions to the *configure* script would let it become just more similar to CMake, which already provides proven and comprehensive mechanisms for dealing with differences between various build host environments and external dependencies. Unfortunately, current versions of OMNeT++ do not support CMake yet.

Therefore, this paper presents various tools, which make CMake aware of OMNeT++'s Network Description (NED) folders and allow to import existing OMNeT++ projects as dependencies. We refer to these dependencies as "legacy projects" in the following. With a CMake based build system it is more convenient to integrate existing implementations of algorithms and suchlike, easing development of OMNeT++ simulation models. Since a simple merger of all dependencies into one project would make complex models unnecessarily huge and hard to keep up-to-date with constantly evolving third-party code, handling these dependencies externally alleviates development in the long term.

The current practice of building OMNeT++ projects is briefly analysed in Section II. Based on this analysis, the components of a joint OMNeT++/CMake build system are presented in Section III. These features are then demonstrated in Section IV using the example of Artery. We conclude with a summary of open challenges and stumbling blocks. Furthermore, suggestions for mainstream CMake support by OMNeT++ are given in Section V.

## II. ANALYSIS

While OMNeT++ itself relies on traditional Makefiles for building its libraries from sources, it provides the Makefile generator *opp_makemake* for creating Makefiles for OMNeT++ projects. Projects like Veins and INET [7] are based upon *opp_makemake*, although they have extended it by individual Python scripts to deal with project specific details. In the case of INET, this script is called *inet_featuretool* and is used for feeding configuration parameters to *opp_makemake* by extracting build information from INET's *.oppfeatures*, *.oppfeaturestate* and *.nedfolders* configuration files. The purpose of this process is to enable or disable INET features. Veins encapsulates the *opp_makemake* invocation in its *configure* script. By default, Veins is built as a shared library and run via OMNeT++'s *opp_run* application. The *configure* script defines the directories which shall be passed to *opp_run* for locating the built shared library and accompanying NED folders. As an option, the location of the INET framework can be given to the script as an argument, so INET modules can be used within a Veins simulation.

In order to facilitate adaptation of existing models and libraries with CMake, no changes to these legacy projects should be required. Albeit their sources are generally available, maintaining changes for CMake compatibility alongside the upstream repositories is an unpleasant task. Furthermore, a generic solution excels individual CMake adaptation code for each legacy project. Thus, additional configuration and maintenance efforts are kept at a bare minimum or avoided completely beforehand.

## III. BUILD SYSTEM DESIGN

There exist three types of dependencies a CMake build system for OMNeT++ projects needs to be aware of. First of all, there are executables, such as the essential C++ compiler as well as *nedtool* (formely *opp_msgc*) bundled with OMNeT++, which is responsible for generating C++ sources and headers from OMNeT++ message descriptions. Secondly, there are static and shared libraries, which can be either plain libraries like Boost or those containing OMNeT++ simulation model code, e.g. the INET shared library. Last but not least, running simulations depends on configuration files like *omnetpp.ini* and NED files accompanying OMNeT++ modules defining their inter-connections and configuration parameters.

### A. CMake fundamentals

A CMake project comprises a platform independent *CMakeLists.txt* file describing the required steps to build targets like executables and libraries. These descriptions make use of commands, variables and properties attached to e.g. targets, directories and source files. One of CMake's cornerstones is for sure its *find_package* mechanism, which supports the integration of a plethora of external dependencies out of the box. The build process of CMake projects involves three phases:

1) *Configuring* the build directory where host-specific information such as locations of dependency libraries and build-specific options like compiler flags are stored. This build configuration can be modified interactively, e.g. with *cmake-gui*.
2) *Generating* out of these information the required files for a native build tool, e.g. a Makefile or IDE project files. In contrast to *CMakeLists.txt*, these files are specific to the build directory and system they are generated for.
3) *Building* the project within the build directory by invoking the native build tool, e.g. GNU Make, or loading the generated build files into the appropriate IDE and launching the build there.

CMake can be extended in various ways, e.g. by creating macros defining custom functionalities building upon CMake's commands. For more details we point to the extensive CMake documentation [1].

All of our subsequently presented CMake extensions and code contributions are available online[1].

---
[1] https://github.com/riebl/artery/releases/tag/opp-summit2015

### B. OMNeT++ simulation environment

Obviously, finding the OMNeT++ environment on the local host is a basic prerequisite for any OMNeT++ project. Unfortunately, neither CMake nor OMNeT++ are aware of each other yet, so there is no official support for CMake's aforementioned *find_package* command to integrate OMNeT++ in the build process. Therefore, we sustain *find_package* by providing the *FindOmnetPP.cmake* script file. It contains heuristics utilised by *find_package* to look for an OMNeT++ installation with its header include directory as well as libraries. If the *omnetpp* executable (the IDE) is within the host's PATH environment variable, the script is capable of detecting the OMNeT++ root directory automatically. OMNeT++ version information as well as several compile definitions are extracted from the therein located *Makefile.inc*. OMNeT++'s libraries are afterwards available as imported targets with appropriate include directories and compile definitions set as interface dependencies, i.e. those information are automatically added to any user target depending on an imported OMNeT++ target library. In contrast to a simple variable pointing at an OMNeT++ library, the employed target mechanism allows to differentiate between release and debug builds, so OMNeT++ libraries with the "d" suffix in their filename are used when a project is built in a debug configuration. Although the simulation model IPv6Suite [8] has used CMake already 10 years ago, the techniques used back then are not on a par with the facilities offered by recent versions of CMake, as e.g. the import of targets inclusive of accompanying target properties such as required header include directories.

### C. Legacy OMNeT++ projects

While these extensions to the *find_package* mechanism are sufficient for building stand-alone OMNeT++ projects with CMake, referencing already existing projects requires further steps. Though adding CMake build files to such legacy projects is feasible, a fast adoption is unlikely and maintenance of two build systems – opp_makemake and CMake – can be a burden. Therefore, a technique to integrate existing OMNeT++ projects like Veins or INET without requiring changes in their respective code bases has a better chance of acceptance.

Fortunately, Makefiles generated by *opp_makemake* comprise the original arguments passed during the generation process, i.e. include directories, compile definitions as well as name, type and location of the produced project binary. These information are processed by the new *opp_cmake* tool, which generates a CMake file wrapping the legacy project's binary and build instructions required for using this legacy binary. The basic process is outlined in Figure 1. At first the legacy project, which is going to be used as a dependency later on, is built in the common way. Usually, a custom Makefile invokes *opp_makemake* with the correct set of arguments. The resulting Makefile includes rules to build the actual legacy binary, which is e.g. a shared library. During the generation phase of CMake, the new *opp_cmake* tool is executed, which parses the original arguments passed to *opp_makemake* from the Makefile and creates a CMake file with import targets for the legacy binary. Finally, this import target file is used via *import_opp_target* for the actual build phase, i.e. the CMake project binary can refer to the legacy binary as an ordinary CMake target. Thus, the project binary can be linked to the legacy binary easily.

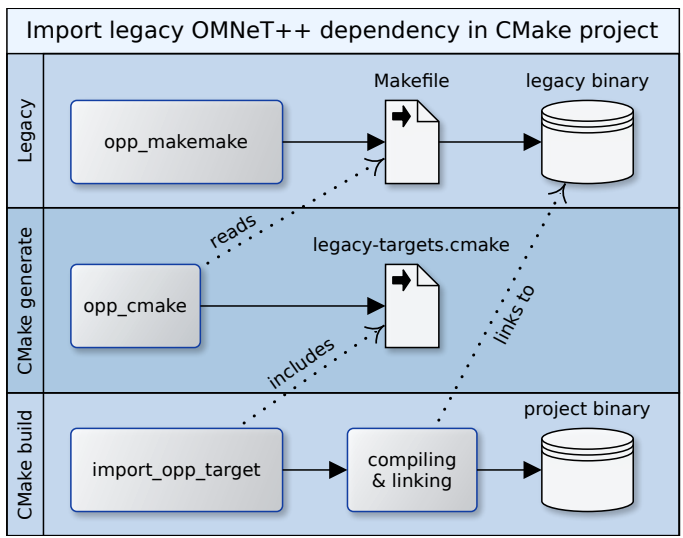

Fig. 1. Import of a legacy project as dependency of a CMake project

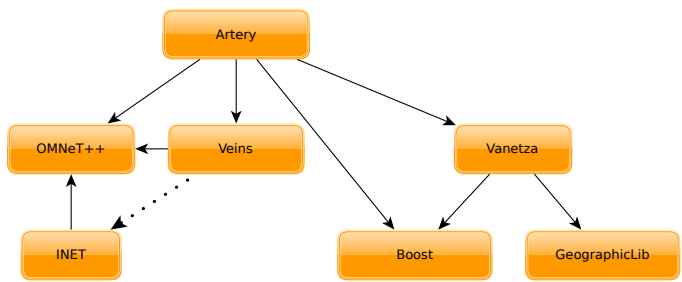

Fig. 2. Dependency graph of Artery

It has to be noted, that a CMake project is only required to call the provided *import_opp_target* macro once. As part of that, the location to the Makefile generated by *opp_makemake* and the name of the import target are passed to the macro. Executing the *opp_cmake* tool as well as loading the import targets are handled by *import_opp_target*. Thus, this mechanism is very easy to use.

So far, *opp_cmake* is not capable to produce CMake targets for legacy projects using "deep includes", i.e. where all project directories are explicitly added to the compiler's header search path. This is not due to a general limitation, but the authors experienced no need for this feature up to now. Future versions of OMNeT++'s *opp_makemake* could also generate a CMake file with import targets alongside the present Makefile, which would make an additional tool like *opp_cmake* obsolete.

*D. NED folders*

A notable addition to OMNeT++ dependencies over conventional C/C++ dependencies are the NED folders containing network description files for each OMNeT++ *Simple Module*. These description files wire up the generic OMNeT++ runtime environment with user-provided C++ code and allow e.g. to set parameters of *Simple Module* objects through the textual simulation configuration. These NED files are not strictly required for the build process itself, but when the compiled simulation is to be executed with the help of *opp_run*. Besides the shared libraries containing the compiled *Simple Modules* it expects also the path to the accompanying NED folders. To account for this runtime requirement, imported targets are attributed with a novel NED_FOLDERS target property, which conveys the list of all NED folders specified in the legacy project's *.nedfolders* file.

By means of the *get_ned_folders* macro, all NED folders of a target and its dependencies can be aggregated recursively. Consequently, no NED folder is missed because every dependent folder is transitively added to the target. This feature enables the creation of run targets within CMake, which start simulations by means of *opp_run*, as the *add_opp_run* macro demonstrates. Alternatively, CMake could also create shell scripts including the NED folders information. CMake's *configure_file* command is predestined for such tasks.

IV. DEMONSTRATION

Practicability of the techniques presented in Section III is demonstrated using the example of the Artery project. The main components involved in this setup and their dependency relations are depicted in Figure 2. Quite obviously, components with custom *Simple Modules* depend on OMNeT++ base classes, just like Artery, INET and Veins do. INET is as of today not strictly required, but might be included implicitly through Veins' *configure* script. Artery, however, might incorporate INET through *import_opp_target* in the future as well, just as it does with Veins too. Additionally, Artery depends on the pure C++ library Vanetza. Since Vanetza's build system is already based on CMake, it can export its targets in its build tree. This works similar to the target import through *import_opp_target*, but here the dependency Vanetza is providing the CMake target file, whereas *opp_cmake* needs to be called by the dependant project. The Boost libraries and headers are integrated via the *find_package* module bundled with CMake. This mechanism is also used for Vanetza's GeographicLib dependency. Consumers of Vanetza, however, do not need to be aware of this dependency because it is transitively added to their dependencies when they just depend on Vanetza. Invoking the "run_example" target generated by CMake starts up the simulation model with an example configuration. If deemed necessary by the build system, changed sources of Artery are automatically rebuilt beforehand.

V. CONCLUSION

Thanks to the presented tools, it is now possible to build an OMNeT++ project using CMake and even incorporate further OMNeT++ projects as dependencies without changing their upstream code base. Despite the advances in handling

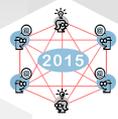

OMNeT++ projects with CMake, however, users still need to take care of installing appropriate versions of all dependencies. This can become slightly challenging in the presence of cross-dependencies, i.e. when multiple components depend on a common third-party library. Mixing several versions of one library into the final library should be avoided. When Veins is built with enabled INET support, other parts, e.g. Artery, should be built with the same INET version to avoid conflicts in the final executable. Same care should be taken regarding other dependencies like Boost.

Oddly enough, shared libraries intended to be executed via *opp_run* are still linked with OMNeT++ libraries. Experiments confirmed this is not required technically. Possibly, this could lead to problems when the shared library is linked with OMNeT++ libraries of a *release* build, whereas *opp_run* refers to *debug* libraries. It needs further investigation if there can arise problems due to this conjuncture.

Instead of creating CMake targets by reverse engineering with *opp_cmake*, *opp_makemake* could create a CMake target file next to the anyway generated Makefile without much effort. This approach would not even interfere with present mechanisms. Since *opp_makemake* is part of OMNeT++, no changes would need to be done to existing projects and therefore enable a smooth transition towards CMake.